\documentclass[useAMS,usenatbib]{mnras}
\usepackage{graphicx, txfonts}
\usepackage{url}
\usepackage{multirow}

\usepackage[usenames,dvipsnames]{color}

\newcommand{\CII}{C\,\textsc{ii}}

\newcommand{\CIII}{C\,\textsc{iii}}
\newcommand{\CIV}{C\,\textsc{iv}}

\newcommand{\FeII}{Fe\,\textsc{ii}}

\newcommand{\HI}{\textrm{H}\,\textsc{i}}

\newcommand{\HII}{\textrm{H}\,\textsc{ii}}

\newcommand{\Lya}{Ly$\alpha$}

\newcommand{\NiII}{Ni\,\textsc{ii}}
\newcommand{\OI}{O\,\textsc{i}}

\newcommand{\SiII}{Si\,\textsc{ii}}
\newcommand{\SiIII}{Si\,\textsc{iii}}
\newcommand{\SiIV}{Si\,\textsc{iv}}

\def\rahr{^{\rm h}}
\def\ramin{^{\rm m}}
\def\rasec{\!\!^{\rm s}}
\def\decdeg{^{\circ}}
\def\decmin{'}
\def\decsec{\!\!''}

\title[The most metal-poor DLA]
{Discovery of the most metal-poor damped Lyman-$\alpha$ system\thanks{
Based on observations collected at the W.M. Keck Observatory
which is operated as a scientific partnership among the California Institute of 
Technology, the University of California and the National Aeronautics and Space 
Administration. The Observatory was made possible by the generous financial
support of the W.M. Keck Foundation.
}}

\author[Cooke et al.]{Ryan J. Cooke$^{1,2,5}$\thanks{email: rcooke@ucolick.org}, 
Max Pettini$^{3}$ and
Charles C. Steidel$^{4}$
\\
$^1$UCO/Lick Observatory, University of California, Santa Cruz, CA 95064, USA\\
$^2$Centre for Extragalactic Astronomy, Department of Physics, Durham University, South Road, Durham DH1 3LE, UK\\
$^3$Institute of Astronomy, Madingley Road, Cambridge, CB3 0HA\\
$^4$California Institute of Technology, MS 249-17, Pasadena, CA 91125, USA\\
$^5$Royal Society University Research Fellow
\\
}

\begin{document}

\date{Accepted . Received ; in original form }
\pagerange{\pageref{firstpage}--\pageref{lastpage}} 
\pubyear{2015}

\maketitle

\label{firstpage}

\begin{abstract}
We report the discovery and analysis of the most metal-poor damped Lyman-$\alpha$ (DLA) system currently known, based on observations made with the Keck HIRES spectrograph. The metal paucity of this system has only permitted the determination of three element abundances:
[C/H]~$=-3.43\pm0.06$,
[O/H]~$=-3.05\pm0.05$, and
[Si/H]~$=-3.21\pm0.05$, as well as an upper limit on the abundance of iron:
[Fe/H]~$\le-2.81$. This DLA is among the most carbon-poor
environment currently known with detectable metals.
By comparing the abundance pattern of this DLA to detailed models of metal-free
nucleosynthesis, we find that the chemistry of the gas is consistent with the yields of
a $20.5~{\rm M_{\odot}}$ metal-free star that ended its life as a core-collapse supernova;
the abundances we measure are inconsistent with the yields of pair-instability supernovae.
Such a tight constraint on the mass of the progenitor Population III star is afforded by the
well-determined C/O ratio, which we show depends almost monotonically on the progenitor
mass when the kinetic energy of the supernova explosion is
$E_{\rm exp}\gtrsim1.5\times10^{51}~{\rm erg}$. We find that the DLA presented here
has just crossed the critical `transition discriminant' threshold, rendering the DLA gas now
suitable for low mass star formation. We also discuss the chemistry of this system
in the context of recent models that suggest some of the most metal-poor DLAs are
the precursors of the `first galaxies', and are the antecedents of the ultra-faint dwarf galaxies.
\end{abstract}

\begin{keywords}
quasars: absorption lines -- ISM: abundances -- stars: Population III -- galaxies: dwarf
\end{keywords}

\section{Introduction}

Almost every astrophysical environment is contaminated by the
nucleosynthesis of stars. To date, only a small handful of pristine
environments have been identified, including a mostly ionized cloud
of gas at redshift $z\simeq3$ which may
be associated with cold flows \citep{FumOmePro11},
and a mostly neutral cloud of gas at redshift $z\simeq7$ attributed
to either a neutral protogalaxy or the intergalactic medium
(\citealt{Sim12}; but see \citealt{BosBec15} for an alternative interpretation).
If pockets of absolutely pristine gas still exist at redshift $z\sim3$, there may
also be some environments that were enriched \textit{exclusively} by the first
generation of stars (also called Population III, or Pop III, stars).

The nature of Pop III stars, in particular their mass spectrum, is still
a matter of debate. Numerical simulations that follow the collapse
of primordial material from cosmological initial conditions originally suggested
that Pop III stars were predominantly very massive, with typical masses
in excess of $100~{\rm M}_{\odot}$ \citep{BroCopLar99,NakUme01,AbeBryNor02}.
Modern calculations suggest a somewhat lower typical mass of the first stars,
and indicate that a small multiple of Pop III stars are formed in a given
minihalo \citep{ClaGloKle08,TurAbeOsh09,StaGreBro10,StaBro13,Hir14,StaBroLee16}.
Although the form of the mass function has not yet been pinned down,
a general conclusion borne out of the above cosmological simulations
is that the initial mass function of the
first stars is top heavy, with most of the total stellar mass
concentrated in stars with masses $M\gtrsim10~{\rm M}_{\odot}$.

In principle, the mass spectrum of the first stars can be inferred
observationally, by identifying the chemical fingerprint of the
elements that were made during the life of a Pop III star. Finding this chemical
signature is a challenging prospect, since a region must be identified
that is \textit{solely} enriched by the first stars. Dedicated surveys to
find putative second generation stars in the Milky Way have
identified several excellent candidates (see \citealt{FreNor15} for a review).
These searches have also
uncovered a striking diversity of chemical abundance patterns,
including many metal-poor stars enhanced with light elements relative
to heavy elements (collectively known as carbon-enhanced
metal-poor [CEMP] stars; for an overview, see \citealt{BeeChr05}),
a star with apparently no iron \citep{Kel14},
and an almost pristine star with roughly solar-scaled chemical
abundances \citep{Caf11}. The abundance patterns of these
stars are successfully reproduced by models of Pop III stellar
nucleosynthesis
\citep{HegWoo10,CooMad14,Ish14,Mar14,TomIwaNom14},
however, there is still some flexibility in the models due
to the uncertain explosion mechanism of core-collapse
supernovae (see e.g. \citealt{Jan12}).

A similar quest has been undertaken to identify the chemical
signature of the first stars among the most metal-poor damped
\Lya\ (DLA) systems \citep{Ern06,Pet08,Pen10,Coo11a,Coo11b,Coo12}.
DLAs are clouds of mostly neutral gas that are observed in
absorption against a bright background source, typically a
quasar (for a general review of DLAs, see \citealt{WolGawPro05}).
DLAs are typically observed at redshift $z\simeq3$ ($\sim2$~Gyr
after the Big Bang) where many of the rest frame ultraviolet
absorption lines of interest are conveniently redshifted into
the optical spectral range. The absorption lines associated with
the most metal-poor DLAs are typically very weak, owing to the
low metal abundance; in such systems, only the most abundant
elements can be reliably measured with current telescope facilities
\citep{Coo13}.

The most metal-poor DLAs are thought to be the
antecedents of the lowest mass galaxies \citep{SalFer12,Web15,CooPetJor15},
as well as cold gas streams being accreted onto galaxies \citep{YuaCen16}.
Such an affiliation marks these DLAs as a promising
environment to measure -- at high redshift -- the chemical
abundance pattern of the earliest generation of stars.

In this paper, we present the discovery of the most metal-poor
DLA currently known, and discuss the chemistry of this
near-pristine environment. In Section~\ref{sec:obs}, we
describe the details of our observations and absorption
line profile fitting. We discuss the chemistry of this
newly discovered DLA in Section~\ref{sec:results}, and
summarise the main conclusions of our work in
Section~\ref{sec:conc}.

\begin{figure*}
  \centering
 {\includegraphics[angle=0,width=170mm]{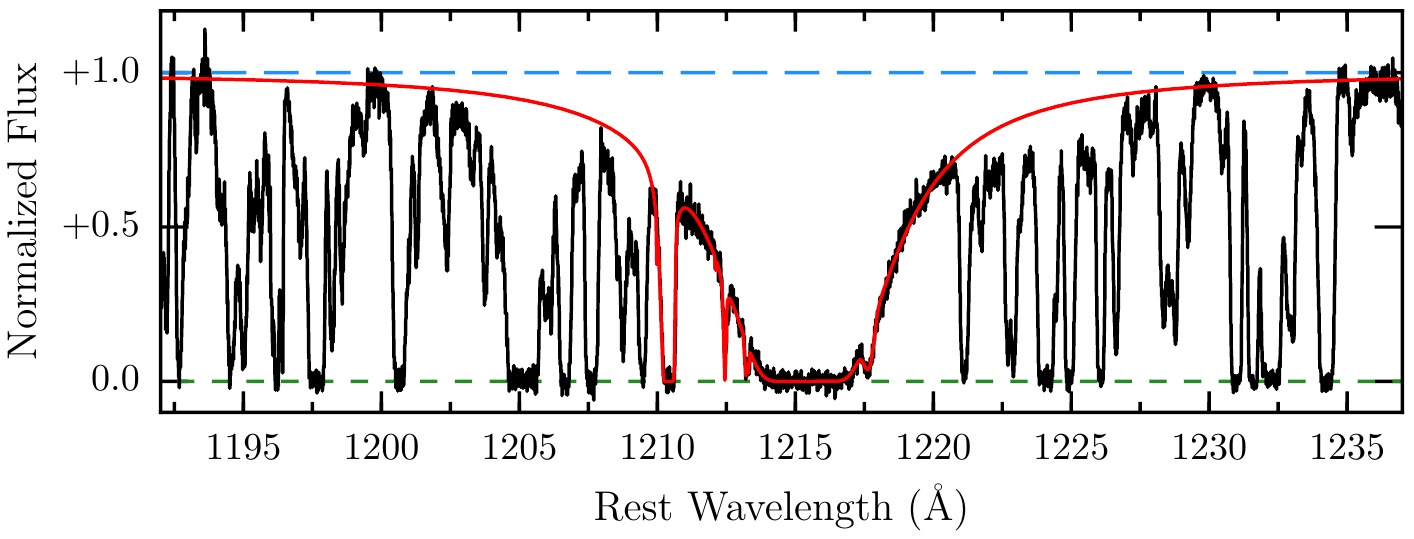}}\\
  \caption{
The \HI\ \Lya\ absorption line of the DLA towards J0903$+$2628 at $z_{\rm abs}=3.07759$ (black histogram) is shown with a Voigt profile model overlaid (red curve) corresponding to a total \HI\ column density of log\,$N$(\HI)/cm$^{-2}$=20.32. Several unrelated blends near the core of the \Lya\ profile are also included in the fit. The short dashed green line indicates the zero level of the data, while the long dashed blue line represents the normalised level of the quasar continuum.
  }
  \label{fig:lya}
\end{figure*}

\section{Observations and Analysis}
\label{sec:obs}

\begin{figure*}
  \centering
 {\includegraphics[angle=0,width=180mm]{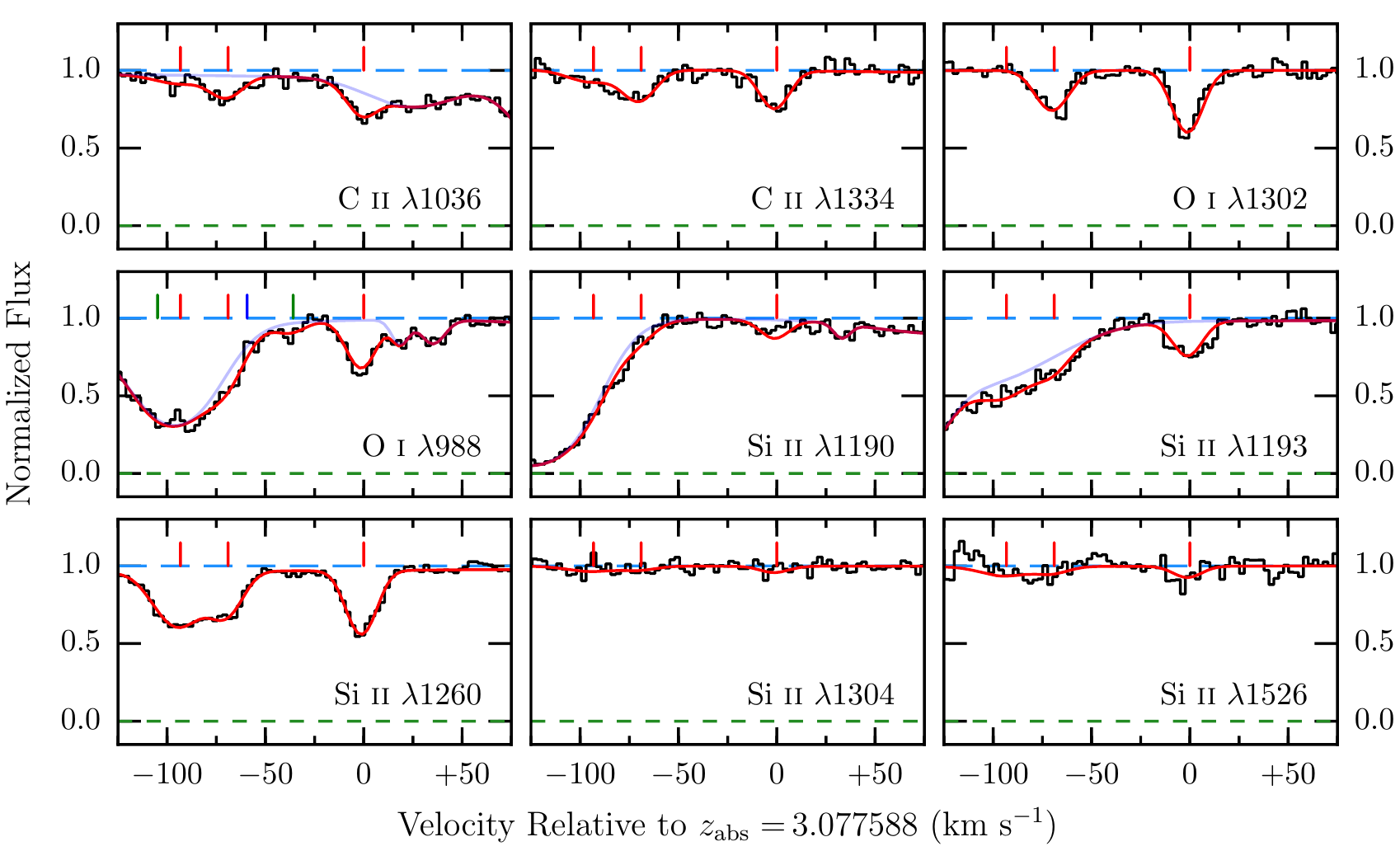}}\\
  \caption{
A selection of metal absorption lines of the DLA towards J0903$+$2628 at $z_{\rm abs}=3.07759$ (black histogram). In each panel, the solid red curve represent the best fitting model profile (including fitted blends). The light purple curves shown in some of the panels represents the model profile of \emph{just} the fitted blends (i.e. absorption that is unrelated to the DLA). The best fitting DLA model profile consists of three components located at relative velocities $v = 0.0$, $-68.9$, and $-93.1$~km~s$^{-1}$ (labelled Components 1, 2 and 3 in the text, respectively), and indicated by the red tick marks above each spectrum. The red, green and blue tick marks in the \OI\,$\lambda988$ panel indicate the model absorption components for the \OI\ triplet, with rest frame wavelengths $\lambda$~=~988.7734\,\AA, 988.6549\,\AA, \rm and~988.5778\,\AA, respectively (note that one/two of the green/blue tick marks are off the plot).
  }
  \label{fig:metals}
\end{figure*}

\begin{table*}
\caption{\textsc{Column densities for each component of the DLA at $z_{\rm abs}=3.07759$ towards J0903$+$2628}}
    \begin{tabular}{@{}lccccccc}
    \hline
   \multicolumn{1}{c}{X}
& \multicolumn{1}{c}{log\,$\epsilon$(X)$_{\odot}\,^{a}$}
& \multicolumn{1}{c}{Component 1}
& \multicolumn{1}{c}{Component 2}
& \multicolumn{1}{c}{Component 3}
& \multicolumn{1}{c}{Total$^{b}$}
& \multicolumn{1}{c}{[X/H]}
& \multicolumn{1}{c}{[X/O]}\\
  \hline
\HI   &  $12.0$  &  \ldots       &  \ldots             &  \ldots  &   $20.32\pm0.05$  &   \ldots  &   \ldots  \\
\CII   &  $8.43$  &  $13.05\pm0.03$       &  $12.99\pm0.04$             &  $12.72\pm0.08$  &   $13.32\pm0.03$ &   $-3.43\pm0.06$  &   $-0.38\pm0.03$  \\
\CIV   &  $8.43$  &  \ldots       &  \ldots             &  \ldots  &   $\le 12.56^{\rm c}$ &   \ldots  &   \ldots  \\
\OI  &  $8.69$  &  $13.74\pm0.02$   &  $13.57\pm0.03$                     &  \ldots  &   $13.96\pm0.02$  &   $-3.05\pm0.05$  &   \ldots  \\
\SiII  &  $7.51$  &  $12.40\pm0.01$  &  $12.23\pm0.03$             &  $12.54\pm0.02$  &   $12.62\pm0.01$  &   $-3.21\pm0.05$  &   $-0.16\pm0.02$  \\
\SiIV   &  $7.51$  &  \ldots       &  \ldots             &  \ldots  &   $\le12.10^{\rm c}$ &   \ldots  &   \ldots  \\
\FeII  &  $7.47$  &  $\le12.76$  &  \ldots             &  \ldots  &   \ldots  &   $\le-2.81^{\rm d}$  &   $\le+0.23^{\rm e}$  \\
  \hline
    \end{tabular}
    \label{tab:coldens}

\begin{flushleft}
\hspace{1.9cm}$^{\rm a}${log\,$\epsilon$(X) = 12 + log\,$N({\rm X})/N({\rm H})$. Solar values are taken from \citet{Asp09}.}\\
\hspace{1.9cm}$^{\rm b}${The total column density only includes the components which are mostly neutral (i.e. Component 1 and 2).}\\
\hspace{1.9cm}$^{\rm c}${The upper limits on $N({\rm \CIV})$ and $N({\rm \SiIV})$ are integrated over the velocity interval $-100\le v/{\rm km~s}^{-1}\le+20$ relative to Component 1.}\\
\hspace{1.9cm}$^{\rm d}${The upper limit on [Fe/H] is based on the [Fe/O] limit of Component 1, combined with the total [O/H] abundance of the DLA.}\\
\hspace{1.9cm}$^{\rm e}${The [Fe/O] abundance is based only on the column densities of Component 1.}\\
\end{flushleft}
\end{table*}

Among the DLAs discovered by the Sloan Digital
Sky Survey \citep{Not09}, we first identified the extremely
metal-poor DLA towards the $z_{\rm em}=3.22$
quasar J0903+2628 (R.A. = $09\rahr03\ramin33.\rasec55$, decl. = $+26\decdeg28\decmin36.\decsec3$; $m_{\rm r}=19.0$) based on the
strong, damped \Lya\ absorption feature at redshift
$z_{\rm abs}=3.076$, combined with the non-detection
of the strongest associated absorption lines of C, O and Si \citep[see][for a description of the technique]{Pet08}.

We observed J0903+2628 with the Keck High Resolution
Spectrometer (HIRES; \citealt{Vog94}) on 2016 March 1, 2, 30
for a total exposure time of $9\times3600$~s, in seeing conditions (full width at half maximum, ${\rm FWHM}\simeq0.8''$)
that were well-matched to the chosen slit width ($0.861''$; decker C1). The nominal instrument resolution of our setup is $R\simeq49,000$, assumed to be a Gaussian profile with a full width at half maximum of $v_{\rm fwhm}=6.1$~km~s$^{-1}$.
We used the blue cross-disperser
to cover the wavelength range 3700--6530\,\AA,
with small gaps near 4600\,\AA\ and 5600\,\AA,
corresponding to the gaps between the detector mosaic.
All frames were binned $2\times2$ during readout.

The data were processed with the \textsc{makee} data reduction
pipeline\footnote{\textsc{makee} is available for download from:\\
http://www.astro.caltech.edu/$\sim$tb/ipac\_staff/tab/makee/index.html}.
This pipeline first subtracts the detector bias level. The locations of the
echelle orders are traced using an exposure of a quartz lamp through
a pinhole decker (HIRES decker D5). The pixel-to-pixel variations and
the blaze function were removed by dividing the science frames by an
exposure of a quartz lamp through the science slit (decker C1).
A one-dimensional spectrum was optimally extracted from each
reduced frame, and the data were finally calibrated to a vacuum
and heliocentric wavelength scale.
The individual spectra were resampled and combined using the \textsc{uves\_popler}
code\footnote{\textsc{uves\_popler} is maintained by Michael T. Murphy, and is
available from the following url:\\ http://astronomy.swin.edu.au/$\sim$mmurphy/UVES\_popler}.
Deviant pixels and ghosts were identified by visual inspection, and masked prior to combination.

The absorption lines were analyzed using the Absorption LIne Software
(\textsc{alis}) package\footnote{\textsc{alis} is available for download from:\\https://github.com/rcooke-ast/ALIS}, which uses the atomic data tabulated by \citet{Mor03}.
Both the quasar continuum emission and the DLA absorption lines were fit simultaneously. The quasar continuum was modeled with a low order Legendre polynomial locally around each absorption line, while each DLA absorption component was modeled with a Voigt profile. The \Lya\ profile of the DLA is presented in Figure~\ref{fig:lya} (black histogram) overlaid with the best fitting Voigt profile (red curve), corresponding to a neutral hydrogen column density of log~$N$(\HI)/cm$^{-2}=20.32\pm0.05$.

We detect metal absorption lines of \CII, \OI\ and \SiII, reproduced in Figure~\ref{fig:metals} together with the best fitting model profiles. The absorption profiles can be described by three components, located in two distinct velocity intervals. The primary absorption component is located at $z_{\rm abs}=3.077588\pm0.000002$, and exhibits a turbulent Doppler parameter of $b_{1}=9.4\pm0.2$~km~s$^{-1}$. The two additional `satellite' components are located at a velocity of $\Delta v_{2}\simeq-68.9$~km~s$^{-1}$ and $\Delta v_{3}\simeq-93.1$~km~s$^{-1}$ relative to the primary component, with Doppler parameters $b_{2}=10.8\pm0.5$~km~s$^{-1}$ and $b_{3}=15.9\pm0.7$~km~s$^{-1}$, respectively. In addition to the turbulent broadening quoted above, we assume that the line profiles of all absorption components are thermally broadened by gas at a kinetic temperature of $T_{\rm kin}=10^{4}$~K \citep[see][]{CooPetJor15}\footnote{We note that the choice of $T_{\rm kin}=10^{4}$~K does not impact the derived column densities, since the thermal Doppler parameter is much narrower than the turbulent Doppler parameter and, in any case, all of the fitted absorption lines are on the linear part of the curve of growth where the derived column density is independent of the Doppler parameter.}. Some of the DLA absorption lines shown in Figure~\ref{fig:metals} are mildly blended with absorption unrelated to the DLA. We simultaneously fit the DLA absorption and the unrelated absorption, and display the model profile of the unrelated blends in Figure~\ref{fig:metals}, represented by the light purple line.

The column densities of each component of the detected ions are listed in Table~\ref{tab:coldens}, and are mostly driven by \CII\,$\lambda1334$, \OI\,$\lambda1302$, and \SiII\,$\lambda1260$. We also list a $3\sigma$ upper limit on the \FeII\ column density, which is based on the non-detection of \FeII\,$\lambda1260$, integrated over the velocity interval of Component 1. The errors on the column density measurements are computed using the diagonal terms of the covariance matrix. We do not detect any absorption originating from the higher ionization absorption lines of \CIV\ and \SiIV. In Table~\ref{tab:coldens} we quote $3\sigma$ upper limits on the column density of these ions. Other high ionization absorption lines of interest (such as \CIII\,$\lambda977$ and \SiIII\,$\lambda1206$) are blended with the \Lya\ forest.

We note that Component 3 is only detected in \SiII\,$\lambda1260$ and is marginally detected in both \CII\ lines; this component is not observed in \OI. Thus, this component may be an unfortunate blend of unrelated
features, not physically associated with the DLA.. However, if this component is associated with the DLA, it probably arises in a region of ionized gas, since no absorption is detected in \OI. For these reasons, we consider only the column density from absorption components 1 and 2 when assessing the relative element composition of this DLA. This ensures that we only consider the mostly neutral gas, where the observed ions are all the dominant ionization stage, thus obviating the need to perform ionization corrections.

\section{Results}
\label{sec:results}

\begin{figure*}
  \centering
 {\includegraphics[angle=0,width=120mm]{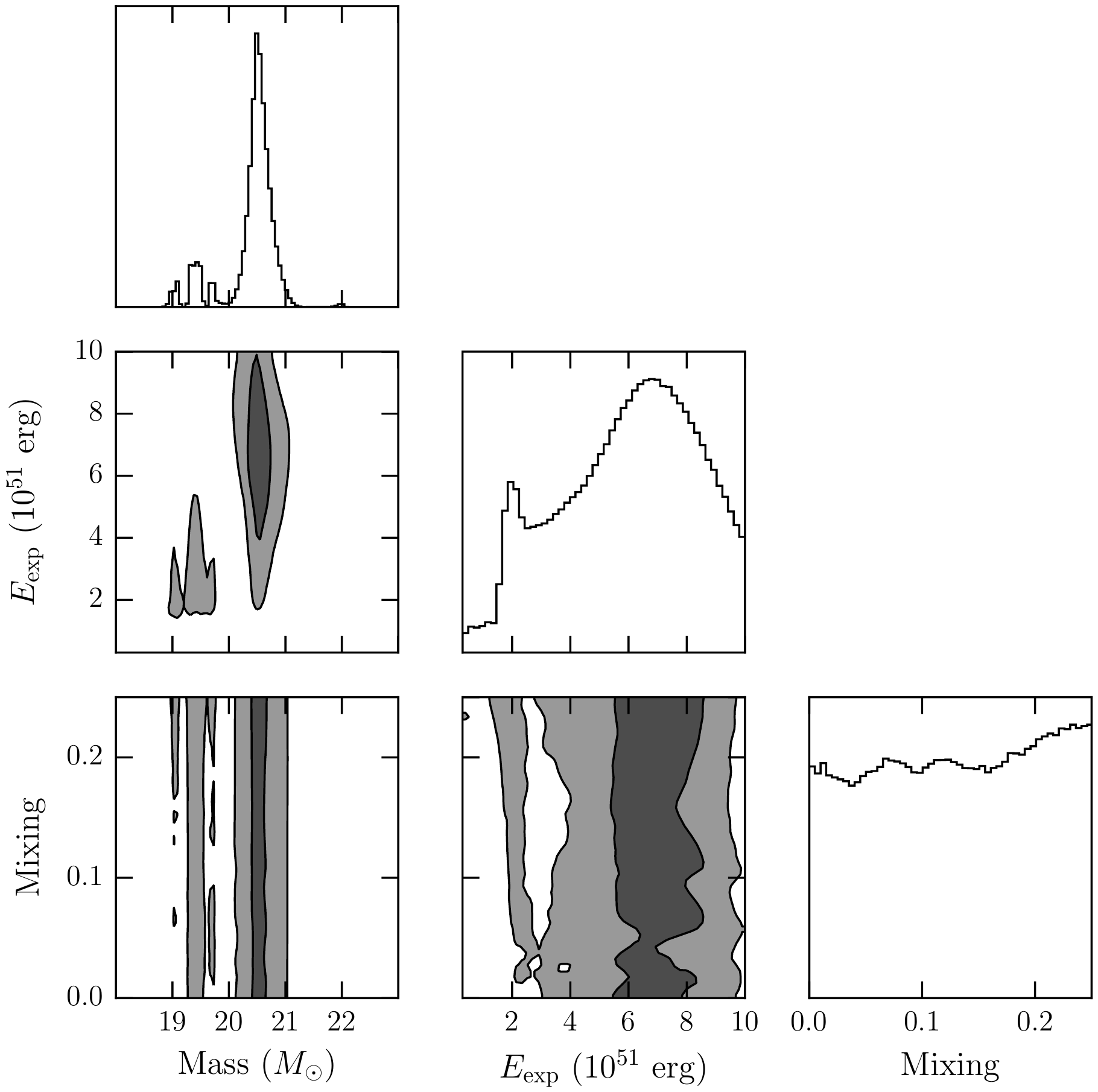}}\\
  \caption{
The observed [C/O] and [Si/O] abundances of the reported DLA have been combined with the \citet{HegWoo10} metal-free nucleosynthesis calculations to estimate the progenitor mass, explosion energy ($E_{\rm exp}$) and stellar mixing parameter of the metal-free star that might have enriched the DLA. The diagonal panels show the marginalised probability of each model parameter shown on the x-axis. The middle left, bottom left, and bottom middle panels show the two dimensional projections, which demonstrate that the model parameters are not degenerate with one another. Dark and light shades enclose the 68 and 95 per cent confidence contours respectively. The non-contiguous regions in the two dimensional projections are due to small changes in the nucleosynthesis of similar mass stars; these changes are caused by slight differences in the location of the various burning shells. The current observations place a strong bound on the mass of the progenitor star, independently of the explosion energy and stellar mixing parameter (see also, Figure~\ref{fig:models}); note that the mixing parameter is unconstrained by the current data. 
  }
  \label{fig:mcmc}
\end{figure*}

\begin{figure}
  \centering
 {\includegraphics[angle=0,width=85mm]{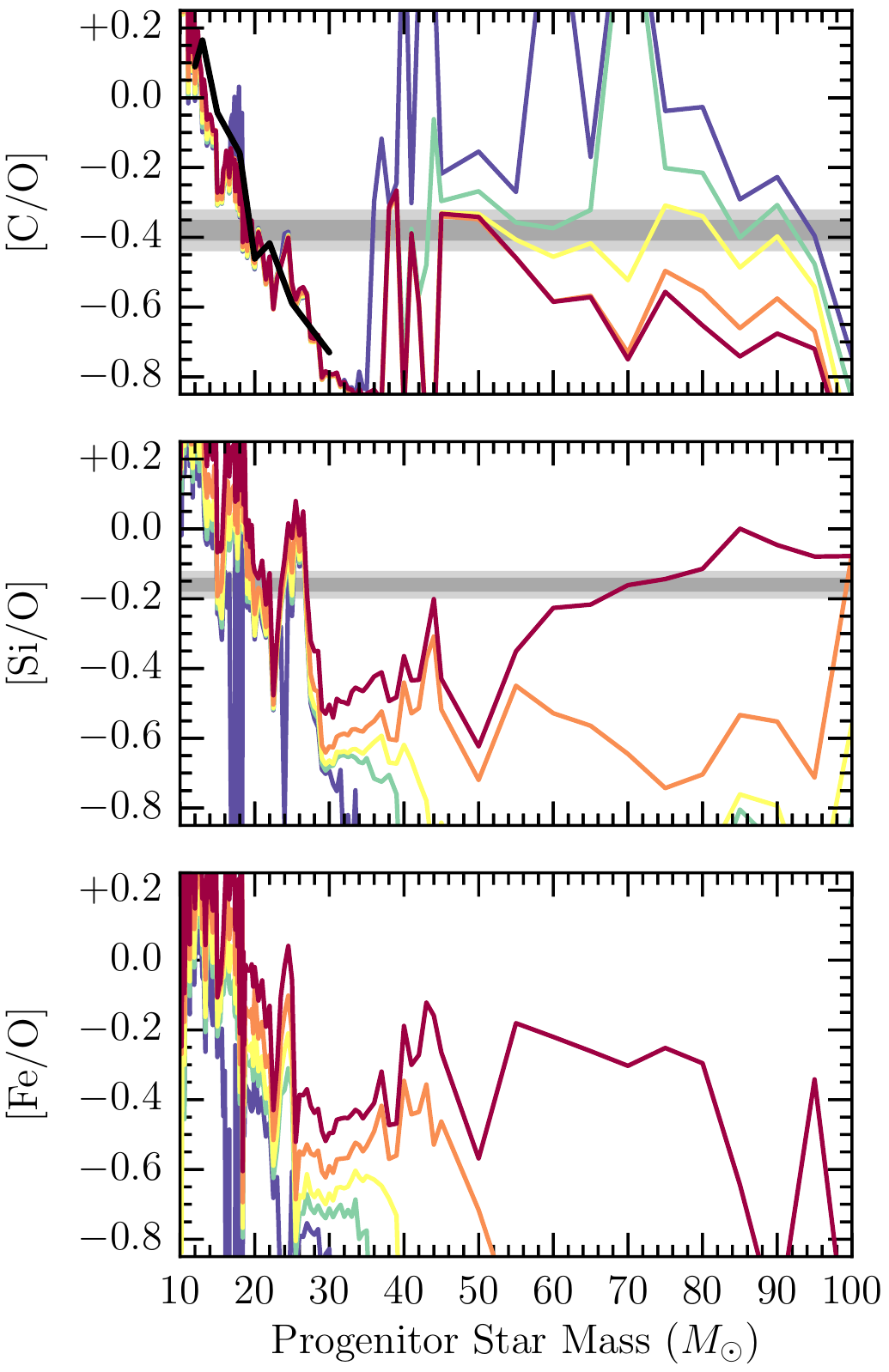}}\\
  \caption{
The calculated relative abundances of [C/O], [Si/O], and [Fe/O] for metal-free stars \citep{HegWoo10} are shown as a function of the progenitor star mass. Each curve is colour-coded by the kinetic energy released by the supernova explosion (${\rm blue \to red}~=~1.8,~2.4,~3,~5,~10~\times~10^{51}~{\rm erg}$). The black curve in the top panel shows the relationship between C/O and progenitor star mass for stars with a metallicity $Z=10^{-4}~Z_{\odot}$ \citep{WooWea95}. The grey shaded regions in the top and middle panels show the abundance measurements of the DLA reported here (dark and light shades represent the 68 and 95 per cent confidence bounds, respectively).
  }
  \label{fig:models}
\end{figure}

\subsection{Metal-poor chemistry}
\label{sec:chemistry}

The DLA reported here is the most metal-poor system currently known, and
has an oxygen abundance ([O/H]~$=-3.05$) a factor of 2 lower than the next most
metal-poor DLA (reported recently by \citealt{Coo16}; [O/H]~$=-2.85$).
Similarly, this system exhibits the lowest C and Si abundance of any known DLA.
Other extremely metal-poor quasar absorption line systems are known with lower
\HI\ column density (known as Lyman limit systems, LLSs), including two systems with no
detectable metals (${\rm [Z/H]}\lesssim-4.0$; \citealt{FumOmePro11}) and two systems
with comparable (lower) metal abundances to the DLA reported here \citep{CriOmeMur16,Leh16}.
Given that the abundance measurements of LLSs require an ionization
correction, DLAs tend to afford somewhat higher precision abundances.
This, in turn, allows for a more informative test of nucleosynthesis models.

The relative element abundances\footnote{In what follows, we assume that the
metals and hydrogen of this DLA are uniformly mixed. This assumption receives
some support from observations of three very metal-poor DLAs
(Q0913$+$072, J1419$+$0829, J1558$-$0031) which exhibit two
absorption components that share a similar metallicity and
abundance pattern (see Table~2 of \citealt{CooPetJor15}; see also, \citealt{Pro03}).}
of this DLA are [C/O]~$=-0.38\pm0.03$ and [Si/O]~$=-0.16\pm0.02$.
These abundances are somewhat lower than the typical
values seen in other very metal-poor DLAs (i.e. DLAs with [Fe/H]~$\le-2.0$),
which have [$\langle$C/O$\rangle$]~$=-0.28\pm0.12$ and
[$\langle$Si/O$\rangle$]~$=-0.08\pm0.10$ \citep{Coo11b}, where the quoted
uncertainty represents the $1\sigma$ spread in the very metal-poor
DLA population. We also find that the [Si/O]
abundance is identical between Components 1 and 2, whereas
the [C/O] abundance of these components differs by 0.11~dex
(a $1.8\sigma$ difference). This difference is still well within the
scatter of all DLA measurements (see Section~\ref{sec:co}).

We first compare the abundance pattern of this DLA to
nucleosynthesis models of very massive metal-free stars
($M\simeq140-260~{\rm M}_{\odot}$) that ended their life
as pair-instability supernovae \citep{HegWoo02}. In these models,
stars with a mass in this range produce element yields of
$-0.66\lesssim{\rm[C/O]}\lesssim-0.60$ and
$+0.20\lesssim{\rm[Si/O]}\lesssim+0.80$,
which are inconsistent with the values we measure.
It thus seems unlikely that
a pair-instability supernova enriched this DLA.

We also compare the abundance pattern of this DLA to nucleosynthesis
models of massive metal-free stars that
ended their lives as type-II core-collapse
supernovae \citep{HegWoo10}. This set of models consists of 120 simulated stars
covering a mass range $M=10-100~{\rm M_{\odot}}$, with a mass resolution
of $\Delta M\gtrsim0.1~{\rm M_{\odot}}$. Since the explosion
mechanism of a type-II supernova is still uncertain, these models
adopt a parameterised `mixing and fallback' scheme. The mixing between
stellar layers during the explosion is achieved by `smoothing' the star over the
mass (i.e. radial) coordinate 4 times with a boxcar filter of a specified
width\footnote{\citet{HegWoo10} adopt this parameterisation as it provides
a good fit to the hard X-ray and optical light curves of SN\,1987A.}; a grid of 14 different widths are
considered by \citet{HegWoo10}, where the mixing width is defined as a fraction of the He core size.
Then, the explosion is simulated as a moving piston that
deposits momentum at a specified mass coordinate. In what follows, we consider
the standard case recommended by \citet{HegWoo10}, which places the piston near
the base of the oxygen burning shell, where the entropy per baryon $S\simeq4\,k_{\rm B}$,
where $k_{\rm B}$ is the Boltzmann constant. The explosion is parameterised
by the kinetic energy of the material that escapes the binding energy of the star,
$E_{\rm exp}$ (hereafter referred to as the explosion energy).
These models consider a grid of 10 values of the explosion energy for each star,
in the range $(0.3-10)\times10^{51}~{\rm erg}$. Thus, this model suite comprises
a total of 16\,800 combinations of the three parameters:
Stellar mass, explosion energy, and mixing width. The ranges of the model parameters
are observationally motivated (see the discussion by \citealt{HegWoo10}), and encompass
all realistic model values.

To find the range of parameters that are an acceptable solution to the abundance
pattern of the DLA reported here, we have linearly interpolated this three-dimensional
space, and conducted a Markov chain Monte Carlo analysis to search for the most likely
set of parameters, using the \textsc{emcee} software \citep{For13}.
In Figure~\ref{fig:mcmc}, we plot the one- and two-dimensional projections
of the samples to identify the covariance between model parameters.
(dark and light shades in the 2D projections represent the 68 and 95
per cent confidence contours, respectively). We find that the progenitor mass
is well-determined, centered on a value $M=20.5~{\rm M_{\odot}}$, while the
supernova explosion energy tends towards the upper end of the
range considered, with a favoured value
$E_{\rm exp} \sim 6$--$8 \times 10^{51}$\,erg.
The mixing parameter is unconstrained.\footnote{Since we only have two measured
relative abundances to fit three model parameters, we are unable to fully constrain
the best-fitting model parameters. The mixing parameter cannot be determined
with the currently available data.}

It may seem surprising that the progenitor mass of the model Population III star
is so well-determined relative to the other parameters of the model. This strong
bound is largely due to the well-determined [C/O] value. The most abundant
isotopes of C and O (i.e. $^{12}$C and $^{16}$O) are primarily formed by
helium burning, with some $^{16}$O resulting from neon burning
\citep[see e.g.][]{WooWea95}. These models suggest that the [C/O] abundance is largely insensitive
to the details of the explosion when $E_{\rm exp}\gtrsim1.5\times10^{51}~{\rm erg}$,
and is almost uniquely dependent on the mass of the progenitor star
(see top panel of Figure~\ref{fig:models}). When the explosion energy
is lower than this value, a higher fraction of $^{16}$O (relative to $^{12}$C)
falls back onto the compact remnant, thereby causing the C/O ratio to
increase \citep[see e.g. Figure 4 of][]{HegWoo10}.
This is matched by a decrease in the Si/O ratio, since more Si falls back
relative to O.

The monotonicity of the relationship between the [C/O] abundance and the progenitor
mass (as well as the invariance of [C/O] with $E_{\rm exp}$) breaks down above
$\sim35~M_{\odot}$, where higher C/O values are recovered by the models.
We note that model stars with a mass above $\sim35~M_{\odot}$ are ruled out by the measured [Si/O]
abundance of the DLA reported here (see middle panel of Figure~\ref{fig:models}).

Given enough data, we
speculate that the strong monotonic dependence of the C/O ratio on
progenitor mass might allow a robust measure of the Pop III initial
mass function. The main uncertainty governing the calculated [C/O]
ratio is the $^{12}$C($\alpha,\gamma)^{16}$O reaction rate that is
used as input into the stellar models; the value of this rate adopted
by \citet{HegWoo10} is consistent with the latest
empirical determination reported by \citet{An16}. However, we caution
that other effects such as rotation \citep{Hir07,Eks08,Jog09}, mass loss
via rotation/winds \citep{MeyMae02}, and the three-dimensional modelling
of metal-free stars \citep{Jog10} may alter the dependence of
C/O on the progenitor mass.

The somewhat weaker bound on $E_{\rm exp}$ shown in Figure~\ref{fig:mcmc} is derived from
the measured [Si/O] ratio (see middle panel of Figure~\ref{fig:models}). We
note that a tighter bound on the explosion energy could be afforded by the [Fe/O]
ratio (bottom panel of Figure~\ref{fig:models}). Unfortunately, the currently
measured upper limit on this ratio ([Fe/O]~$\le+0.23$), is unable to provide
an informative bound on $E_{\rm exp}$. Since the [Fe/O] abundance
is also highly sensitive to the chosen mixing parameter, a measurement of
the [Ni/Fe] abundance is needed to break the degeneracy (see \citealt{Coo13}).
These measurements might become possible with the next generation of
30+\,m telescope facilities (see Section~\ref{sec:ufds}).

Finally, we have only two relative element abundances with which to
constrain the nucleosynthesis models, and we are unable at this stage
to rule out the possibility that this system may be contaminated by other
forms of nucleosynthesis, including Population II core-collapse
supernovae. At a redshift of $z_{\rm abs}=3.07759$, the Universe
has aged by $\simeq1.5$\,Gyr after the epoch of the first stars
($z\simeq10-15$; e.g. \citealt{Mai10}), and there is enough time
for a second generation of stars to have already operated in this
DLA. It is nevertheless encouraging that the chemistry
of this one DLA is in good agreement with the chemistry of absorption
line systems at $z_{\rm abs}\simeq6$ \citep{Bec12}, which are
captured only a few hundred Myr after the putative epoch of the
first stars. It is also worth noting that the low metallicity of this DLA
is consistent with the metallicity regime expected for gas that is
enriched solely by metal-free Population III stars
\citep{SmiSig07,BroYos11,Wis12,CooMad14,Web15,Rit16}.

\subsection{The antecedents of metal-poor stars and the ultra-faint dwarf galaxies}
\label{sec:ufds}

According to the Stellar Abundances for Galactic Archaeology (SAGA) database\footnote{The SAGA database is maintained by Takuma Suda, Yutaka Katsuta, and Shimako Yamada, and is available at the following address:\\
http://sagadatabase.jp/wiki/doku.php} \citep{Sud08}, the only star currently known with a lower [C/H] abundance than the DLA reported here is the Leo star \citep{Caf11}. The remaining stars in the SAGA database are either red giant branch stars\footnote{As a star ascends the red giant branch (RGB), carbon that has been processed through the CN cycle is mixed to the stellar surface, reducing the surface abundance of C \citep[e.g.][]{Gra00}. As a result, the [C/H] abundance that an upper RGB star was born with can only be recovered by applying a correction based on stellar evolution models \citep{Pla14}. After correcting for this effect, \citet{Pla14} find $\sim10$ RGB stars that exhibit a corrected [C/H] value that is lower than the DLA we report here.} or display a higher abundance of [C/H]. The star with the closest abundances to the DLA we report here is SDSS~J0259$+$0057 \citep{Aok13},
which has a carbon abundance [C/H]~$=-3.33\pm0.22$, and is not carbon-enhanced ([C/Fe]~$=-0.02\pm0.22$).

If the DLA we consider here has a chemical abundance pattern that is similar to the Milky Way halo star SDSS~J0259$+$0057, we estimate that the Fe abundance of the DLA would be ${\rm [Fe/H]}\simeq-3.4$. The Fe abundance of the DLA can also be estimated by subtracting the typical Si/Fe abundance of very metal-poor DLAs ([Si/Fe]~$=+0.32\pm0.09$; \citealt{Coo11b}) from the observed [Si/H] abundance, which would give ${\rm [Fe/H]}\simeq-3.5$. We note, however, that the [Fe/H] abundance of this DLA may be considerably lower than these estimates; there exist some Milky Way halo stars with slightly higher [C/H] that exhibit a strong overabundance of [C/Fe], collectively known as CEMP stars \citep[see e.g.][]{BeeChr05}.
Measuring the detailed chemical abundance pattern of this DLA (including the
abundances of N, Mg, Al, and Fe) will become feasible with the next
generation of 30+~m telescope facilities. Assuming that the
abundances of these elements are in the same proportions as a typical very
metal-poor DLA (see Table~13 of \citealt{Coo11b}), this goal could be achieved
with a spectrum of ${\rm S/N}~\simeq~150$.\footnote{The next
detectable elements include S and Ni, which would require a
${\rm S/N}~\simeq~600$ and $2000$, respectively, to detect
an individual absorption line at $5\sigma$ confidence.
Stacking the available \NiII\ lines \citep[see][]{Coo13} could
reduce this requirement to a ${\rm S/N}~\simeq~1000$.}

As stated above, there is only one star currently known that is more
carbon-poor than the DLA reported here. We now
speculate how many DLAs might need to be observed before
a DLA with an abundance as low as the Leo star may be found. The SAGA
database lists 7 stars with a carbon abundance [C/H]~$\le-3.0$,
including the Leo star which has an abundance [C/H]~$\le-3.8$.
There are currently four DLAs with a carbon abundance
[C/H]~$\le-3.0$ (see Section~\ref{sec:co} and Table~\ref{tab:co});
thus, doubling the number of DLAs in this extremely metal-poor
regime may be sufficient to uncover a DLA of similar metallicity
to the Leo star.

\begin{figure*}
  \centering
 {\includegraphics[angle=0,width=140mm]{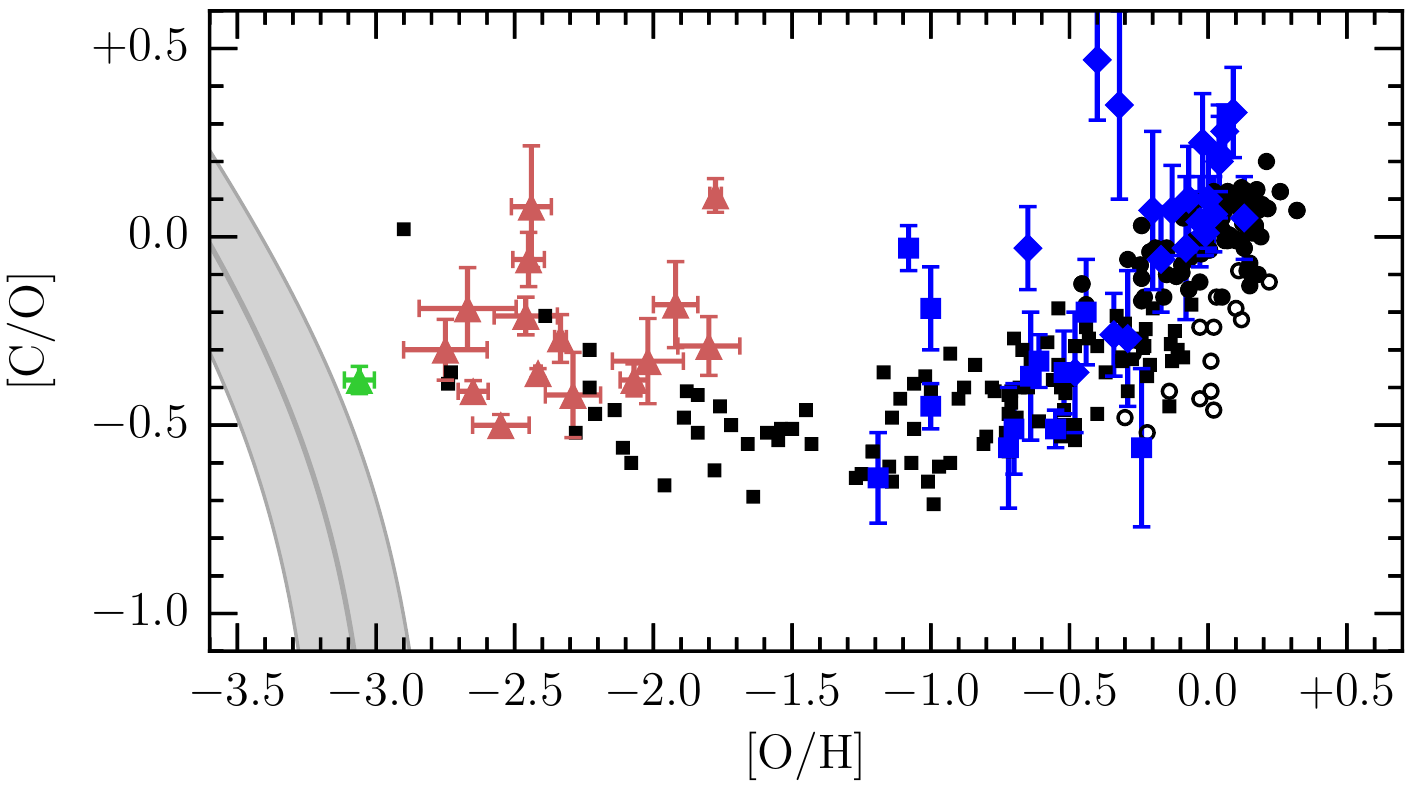}}\\
  \caption{
The chemical evolution of [C/O] is shown for various astrophysical environments. The black symbols represent stars that are kinematically associated with the thin disc (solid circles), thick disc (open circles), and halo stars (squares). The blue symbols are for \HII\ regions, measured from optical recombination lines (diamonds) or ultraviolet collisionally excited lines (squares). The red triangles show literature measurements of [C/O] and [O/H] of metal-poor DLAs (see Table~\ref{tab:co}). The green triangle is the new measurement reported herein. The grey band represents the `transition discriminant' proposed by \citet{FreJohBro07}, where the width of the band indicates the uncertainty of this zone. Gas clouds to the left of this zone are not expected to form a low mass (i.e. long-lived) generation of stars.
See the text for the references to all plotted datasets.
  }
  \label{fig:co}
\end{figure*}

Given the metal paucity of the DLA reported here, it is
conceivable that just a \textit{single} (Population III) star
is responsible for the metal enrichment. Indeed, hydrodynamic
models that trace the internal feedback\footnote{Including
both radiative feedback from massive stars and the
subsequent supernova feedback.} within idealized
dwarf galaxies, do entertain this idea
\citep{Web15,Bla15}. In this scenario, the DLA reported here would be
considered an immediate progenitor of a ``first galaxy'' that is poised to
form its first generation of low mass stars (see also, Section~\ref{sec:co}).
This scenario receives some support from abundance measurements
of stars in the lowest luminosity and most metal-poor
galaxies in the Local Group -- the so-called ultra-faint dwarf galaxies
(UFDs; \citealt{Kir08,Nor10,Sim11,Lai11,Var13,Bro14}); the presumed
[Fe/H] abundance of the DLA reported here ([Fe/H]~$\simeq-3.5$) is
comparable to the \emph{most} Fe-poor stars in the UFDs.\footnote{We
note that a direct comparison between the [Si/H] abundance of this DLA and
the stars in UFDs is not possible, since [Si/H] has not been measured
for the \textit{most} Fe-poor stars in UFDs \citep{Var13}.}

\subsection{Metallicity evolution of the C/O ratio}
\label{sec:co}

\begin{table}
\begin{center}
    \caption{\textsc{Compilation of DLA C/O and O/H measurements}}
    \hspace{-0.6cm}\begin{tabular}{@{}lcccc}
    \hline
   \multicolumn{1}{c}{QSO}
& \multicolumn{1}{c}{$z_{\rm abs}$}
& \multicolumn{1}{c}{[C/O]}
& \multicolumn{1}{c}{[O/H]}
& \multicolumn{1}{c}{ref$^{\rm a}$}\\
  \hline
J0035$-$0918  &  2.34010  &  $0.08\pm0.16$  &  $-2.44\pm0.07$   &  1,2  \\
HS0105$+$1619  &  2.53651  &  $0.110\pm0.045$  &  $-1.776\pm0.021$   &  3  \\
J0140$-$0839  &  3.69660  &  $-0.30\pm0.08$  &  $-2.75\pm0.15$   &  4  \\
J0311$-$1722  &  3.73400  &  $-0.42\pm0.11$  &  $-2.29\pm0.10$   &  5  \\
J0903$+$2628  & 3.07759   &  $-0.38\pm0.03$  &  $-3.05\pm0.05$   &  6  \\
Q0913$+$072  &  2.61829  &  $-0.36\pm0.012$  &  $-2.416\pm0.011$   &  3, 7  \\
J0953$-$0504  &  4.20287  &  $-0.50\pm0.03$  &  $-2.55\pm0.10$   &  2  \\
J1001$+$0343  &  3.07841  &  $-0.41\pm0.03$  &  $-2.65\pm0.05$   &  5  \\
J1016$+$4040  &  2.81633  &  $-0.21\pm0.05$  &  $-2.46\pm0.11$   &  7  \\
Q1111$+$1332  &  2.27094  &  $-0.18\pm0.11$  &  $-1.92\pm0.08$   &  1  \\
Q1202$+$3235  &  4.97700  &  $-0.33\pm0.11$  &  $-2.02\pm0.13$   & 8   \\
J1337$+$3153  &  3.16768  &  $-0.19\pm0.11$  &  $-2.67\pm0.18$   &  9  \\
J1358$+$6522  &  3.06730  &  $-0.27\pm0.06$  &  $-2.335\pm0.02$   &  3, 10  \\
J1558$+$4053  &  2.55332  &  $-0.06\pm0.07$  &  $-2.45\pm0.06$   &  7  \\
Q2206$-$199  &  2.07624  &  $-0.38\pm0.04$  &  $-2.07\pm0.05$   &  7  \\
J2155$+$1358  &  4.21244  &  $-0.29\pm0.08$  &  $-1.80\pm0.11$   &  11  \\
  \hline
    \end{tabular}
    \label{tab:co}
\end{center}

$^{\rm a}${1: \citet{CooPetJor15};
2: \citet{Dut14};
3: \citet{Coo14};
4: \citet{Ell10};
5: \citet{Coo11b};
6: This work;
7: \citet{Pet08};
8: \citet{Mor16};
9: \citet{Sri10};
10: \citet{Coo12};
11: \citet{Des03}.
}\\
\end{table}

We now extend this discussion to the metallicity evolution of
the C/O ratio. In Figure~\ref{fig:co}, we present the latest
complement of C/O and O/H measurements available in the
literature. These include abundance measurements of stars
\citep{BenFel06,Fab09,Nis14} in the Milky Way thin and thick disc
(solid and open black circles respectively)
and in the Milky Way halo (black squares).
To this we add C/O measurements of local \HII\ regions
\citep{DufShiTal82,Gar95,Kur99,Est02,Est09,Est14,GarEst07,Lop07,Ber16},
using either optical recombination lines or ultraviolet collisionally excited lines
(blue diamonds and squares respectively).\footnote{We apply a correction of
$+0.24$~dex to the [O/H] abundance measured using collisionally excited
lines \citep{Est14,Ste16}.} Finally, we overplot measurements of C/O and O/H
in the lowest metallicity DLAs (red triangles) and
in the new DLA reported here (green triangle).
For convenience, in Table~\ref{tab:co} we provide all of the DLA
measurements used in this work, scaled to our adopted solar
abundance scale \citep[][see also the second column of
Table~\ref{tab:coldens}]{Asp09}.

All of the observations displayed in Figure~\ref{fig:co} are generally in good
mutual agreement, regardless of the different environments probed
and the different techniques used. There are, however, a few DLAs and \HII\ regions that exhibit
somewhat enhanced C/O values relative to the locus defined by
Milky Way stars. Similar enhancements have also been seen in
the [C/$\alpha$] ratios measured in some LLSs (\citealt{Leh16}; see also, \citealt{Leh13}).
The origin of such enhancements is unclear at present, but could
be related to the production of C by intermediate mass stars \citep{Sha16}.

As discussed previously in the literature
\citep{Hen00,Car00,Ake04,Ces09,Rom10}, the increase of [C/O] when
[O/H]~$\gtrsim-1.0$ is thought to be due to the metallicity dependent winds
of massive stars combined with the delayed release of C from
low and intermediate mass stars. The upturn of [C/O] when
[O/H]~$\lesssim-1.5$, on the other hand, is still open to debate.
\citet[][see also, \citealt{CarPei11}]{Ake04} propose that the increased C/O values seen at
low O/H is the result of an increased carbon yield from
metal-free Population III stars. This scenario has previously
been used to explain the lowest metallicity population of
CEMP stars \citep{UmeNom03,Rya05,CooMad14}.

In addition to the C/O upturn at low O/H,
we also draw attention to the scatter in the
DLA [C/O] measurements. With the improved
statistics of the data collected in Table 2,
we now see that the DLA values of [C/O]
at a given [O/H] differ by more than their errors,
suggesting an intrinsic dispersion in these
elements relative abundances that was not
immediately apparent in our earlier studies
of carbon and oxygen in very metal-poor DLAs
\citep{Pet08,Coo11b}. The total range in [C/O]
exhibited by DLAs with [O/H]~$\lesssim-2.0$
spans a factor of $\sim4$, from
[C/O]~$\simeq-0.5$ to $\simeq+0.1$.
There are several possibilities that could explain
such a scatter. As shown in the top panel of
Figure~\ref{fig:models} (see also, Section~\ref{sec:chemistry}),
the [C/O] ratio
is sensitive to the mass of the stars that enriched the
DLAs. If the first stars formed in isolation
or in small multiples,
as in currently favoured scenarios
\citep{ClaGloKle08,TurAbeOsh09,StaGreBro10,StaBro13,Hir14,StaBroLee16},
the intrinsic C/O scatter could be explained if these DLAs were
enriched by a single or a small multiple of Pop III stars. The predicted
mass range of the stars responsible for the enrichment of these metal-poor
DLAs is $10\lesssim M/{\rm M_{\odot}}\lesssim25$, with a bias towards higher
mass stars. An alternative possibility, suggested recently by \citet{Sha16},
is that the C/O scatter is due to enrichment by a combination of massive
and intermediate mass (Population II) stars; the high C/O values are the result
of preferential enrichment by intermediate mass stars while the low C/O values
are primarily due to massive stars. The observed C/O scatter is the result of
poor mixing between these two channels of carbon production.

We point out that massive Population II stars alone are less likely to produce the
observed C/O scatter than Population III stars, even though the relationship between
C/O and progenitor
star mass is nearly identical for Population III stars and stars with progenitor metallicity
$Z=10^{-4}~Z_{\odot}$ (see the black curve in the top panel of
Figure~\ref{fig:models}). This is because the stellar initial
mass function of Population II stars is more fully sampled, unlike the sparsely
sampled initial mass function of Population III stars, as discussed above.
Likewise, we suggest that the scatter of C/O measurements in the most
metal-poor DLAs might provide a measure of the stellar initial mass function
and multiplicity of the first stars. Measuring the intrinsic scatter of C/O in
almost pristine environments should be considered a key goal of future
observations.

In Figure~\ref{fig:co}, we also overlay the `transition discriminant'
\citep[][see also \citealt{BroLoe03}]{FreJohBro07},
shown by the solid grey shaded region, which marks the transition from
a predominantly high mass generation of (Population III) stars to a low
mass generation of (Population II) stars. This criterion is based on the
fine structure cooling lines of [\CII] and [\OI]. Once a critical abundance
of C and O is reached in a cloud of cold neutral gas (i.e. to the right of
the grey band shown in Figure~\ref{fig:co}), the gas
is able to fragment to low mass scales and produce a long-lived
generation of stars. In principle, DLAs are not restricted to the right
of this region; a DLA that overlaps this region would be an ideal
environment to empirically study the transition from
Population III to Population II star formation.
The DLA reported here lies very close to this region,\footnote{The DLA
reported here exhibits a value of the transition discriminant, ${D_{\rm trans}=-3.19\pm0.04}$,
which lies just above the critical value calculated
by \citet{FreJohBro07}, $D_{\rm trans,crit}=-3.5\pm0.2$.} and appears
to have just crossed the threshold for low mass star formation.

Finally, given that: (1) The gas in DLAs is mostly neutral (i.e. conducive to star
formation); (2) this particular DLA has just crossed into a chemical regime where
a generation of long-lived stars is expected to form; and (3) the presumed [Fe/H]
abundance of the DLA reported here is consistent with the \textit{most} Fe-poor stars
seen in UFDs, we suggest that this DLA could be representative of
one of the antecedents of the UFD galaxy population, as discussed
previously in Section~\ref{sec:ufds}.


\section{Summary and Conclusions}
\label{sec:conc}

We report the discovery of the most metal-poor damped Lyman-$\alpha$ system
currently known, located at a redshift $z_{\rm abs}=3.07759$, based on observations
of the quasar J0903$+$2628 taken with the Keck HIRES spectrograph.
This DLA was identified as part of our ongoing programme to search for
the nucleosynthetic imprints of the first stars in neutral gas at high
redshift. We draw the following main conclusions:\\

\noindent ~~(i) The extremely low metallicity of this DLA has allowed
only the most abundant chemical elements to be detected. Based on
Voigt profile fitting, we deduce the following abundances:
[C/H]~$=-3.43\pm0.06$,
[O/H]~$=-3.05\pm0.05$, and
[Si/H]~$=-3.21\pm0.05$, and an upper limit on the iron abundance of
[Fe/H]~$\le-2.81$. This DLA is a factor of $\sim2$ more oxygen-poor
than the next most metal-poor DLA known (\citealt{Coo16}; [O/H]~$=-2.85$).

\smallskip

\noindent ~~(ii) This DLA exhibits an abundance of carbon that is lower than
any metal-poor star currently known, other than the Leo star \citep{Caf11}.
The low metallicity is consistent with the view that this DLA may have
been enriched solely by the products of a first generation of stars.

\smallskip

\noindent ~~(iii) We have compared the relative chemical abundances of
this DLA to nucleosynthesis calculations of massive metal-free stars that ended
their lives as core-collapse supernovae. We show that the C/O yield of
massive metal-free stars is a strong (almost monotonic) function of the
progenitor mass, and is virtually independent of the other parameters
in the nucleosynthesis modelling when the explosion energy
is $E_{\rm exp}\gtrsim1.5\times10^{51}~{\rm erg}$.
Using these models, we estimate that the DLA was enriched by a star
of mass $M\simeq20.5~{\rm M}_{\odot}$. We also rule out the possibility
that the enrichment was due to a star that ended its life as a pair-instability
supernova.

\smallskip

\noindent ~~(iv) The carbon and oxygen abundances of this DLA yield a
measure of the transition discriminant, $D_{\rm trans}=-3.19\pm0.04$,
which is just over the critical threshold for low mass star formation
$D_{\rm trans,crit}=-3.5\pm0.2$ \citep{FreJohBro07}.
Given the large column density of neutral gas hosted by the DLA, which
may be conducive to star formation, we propose that this environment might
represent an immediate precursor to the formation of a `first galaxy' (i.e. a dark
matter halo that has not yet formed a generation of long-lived, low mass stars).

\smallskip

\noindent ~~(v) We compile all available literature determinations
of the C/O abundance in very metal-poor DLAs.
The scatter of these measurements is larger
than the measurement uncertainties, indicating
that there is an intrinsic dispersion in the
population. Given the sensitivity of the
C/O ratio to progenitor mass, we propose that the scatter can
be explained if a single or small
multiple of metal-free stars is responsible for
the enrichment of extremely metal-poor DLAs.
We also propose that the distribution of C/O measurements
of extremely metal-poor DLAs can be used to determine both
the stellar initial mass function and the multiplicity of the
first stars.

\smallskip

\noindent ~~(vi) Recent numerical models have also suggested that the
most metal-poor DLAs may have been enriched by just a single
massive star, and could have similar properties to the progenitors
of the Milky Way ultra-faint dwarf galaxies \citep{SalFer12,Web15}.
We note that the presumed [Fe/H] of the DLA reported here is consistent
with the \textit{most} Fe-poor stars found in the UFD galaxies. A key goal
of future work will be to pin down the iron abundance of the DLA reported
here and of other extremely metal-poor DLAs.

\smallskip

Our work highlights the importance of identifying and measuring the
detailed chemical abundance patterns of the most metal-poor DLAs.
The high precision and accuracy that can be achieved for the
relative chemical abundances ($\lesssim0.05$~dex), allow
for a strong and informative test of nucleosynthesis models of metal-free
stars. In particular, the C/O ratio offers a sensitive conversion
to the progenitor mass of the metal-free stars that may have
enriched the most metal-poor DLAs. With future telescope
facilities, it may also be possible to measure the [Ni/Fe] ratio
to comparable precision, thereby allowing a strong bound
on the supernova explosion energy \citep{Coo13}. Finally, we
estimate that by merely doubling the number of DLAs with a
carbon abundance [C/H]~$\le-3.0$ (from 4 to 8 DLAs), it is
statistically possible to find a DLA that is as metal-poor as
the Leo star. Furthermore, the future prospect of finding extremely
metal-poor (and possibly, pristine) DLAs at high redshift is
very promising; the steep slope of the quasar luminosity
function combined with the collecting area of the 30+\,m
telescopes, will ensure access to $\sim100$ times more
quasars than those that are within reach of current facilities
(see the discussion by \citealt{Coo16}).

\section*{Acknowledgements}
We are grateful to 
the staff astronomers at Keck 
Observatory for their assistance 
with the observations.
We thank the anonymous referee for their prompt review,
and for offering several helpful suggestions that improved
the presentation of this paper.
During this work, R.~J.~C. was supported by a Royal Society
University Research Fellowship, and by NASA through
Hubble Fellowship grant HST-HF-51338.001-A, awarded by the
Space Telescope Science Institute, which is operated by the
Association of Universities for Research in Astronomy, Inc.,
for NASA, under contract NAS5- 26555.
CCS has been supported by grant AST-1313472 from the U.S. NSF.
This research was also supported by a NASA Keck PI Data Award,
administered by the NASA Exoplanet Science Institute.
Data presented herein were obtained at the W. M. Keck Observatory
from telescope time partially allocated to the National Aeronautics and Space
Administration through the agency's scientific partnership with the
California Institute of Technology and the University of California.
Our work made use of the \textsc{matplotlib} \citep{Hun07}, \textsc{emcee} \citep{For13},
and \textsc{corner} \citep{For16} python packages, which we gratefully acknowledge.
The Observatory was made possible by the generous financial
support of the W. M. Keck Foundation. We thank the Hawaiian
people for the opportunity to observe from Mauna Kea;
without their hospitality, this work would not have been possible.





\label{lastpage}

\end{document}